\documentclass{article}

\usepackage{arxiv}
\usepackage[authoryear]{natbib}
\usepackage[utf8]{inputenc} 
\usepackage[T1]{fontenc}    
\usepackage{hyperref}       
\usepackage{url}            
\usepackage{booktabs}       
\usepackage{amsfonts}       
\usepackage{nicefrac}       
\usepackage{microtype}      
\usepackage{lipsum}		
\usepackage{graphicx}
\usepackage{natbib}
\usepackage{doi}
\usepackage{graphicx}%
\usepackage{multirow}%
\usepackage{amsmath,amssymb,amsfonts}%
\usepackage{amsthm}%
\usepackage{mathrsfs}%
\usepackage[title]{appendix}%
\usepackage{xcolor}%
\usepackage{textcomp}%
\usepackage{manyfoot}%
\usepackage{booktabs}%
\usepackage{algorithm}%
\usepackage{algorithmicx}%
\usepackage{algpseudocode}%
\usepackage{listings}%
\usepackage{graphicx}
\usepackage{longtable}
\usepackage{array}
\usepackage{booktabs}
\usepackage{multirow}
\usepackage{adjustbox}
\usepackage{xcolor}
\usepackage{url}
\usepackage{hyperref}
\usepackage{algpseudocode}
\usepackage{float}

\definecolor{codegreen}{rgb}{0,0.6,0}
\definecolor{codegray}{rgb}{0.5,0.5,0.5}
\definecolor{codepurple}{rgb}{0.58,0,0.82}
\definecolor{backcolour}{rgb}{0.95,0.95,0.92}

\lstdefinestyle{mystyle}{
    backgroundcolor=\color{backcolour},   
    commentstyle=\color{codegreen},
    keywordstyle=\color{magenta},
    numberstyle=\tiny\color{codegray},
    stringstyle=\color{codepurple},
    basicstyle=\ttfamily\footnotesize,
    breakatwhitespace=false,         
    breaklines=true,                 
    captionpos=b,                    
    keepspaces=true,                 
    numbers=left,                    
    numbersep=5pt,                  
    showspaces=false,                
    showstringspaces=false,
    showtabs=false,                  
    tabsize=2
}

\title{Bug Classification in Quantum Software: A Rule-Based Framework and Its Evaluation}

\author{{\hspace{1mm}Mir Mohammad Yousuf} \\
	Department of Information Technology\\
	National Institute of Technology Srinagar\\
	J\&K, India, 190006 \\
	\texttt{yousuf\_2022phaite006@nitsri.ac.in} \\
	\And
	{\hspace{1mm}Shabir Ahmad Sofi} \\
	Department of Information Technology\\
	National Institute of Technology Srinagar\\
	J\&K, India, 190006 \\
	\texttt{shabir@nitsri.ac.in} \\
}

\begin{document}
\maketitle

\begin{abstract}
	Accurate classification of software bugs is essential for improving software quality. This paper presents a rule-based automated framework for classifying issues in quantum software repositories by bug type, category, severity, and impacted quality attributes, with additional focus on quantum-specific bug types. The framework applies keyword and heuristic-based techniques tailored to quantum computing. To assess its reliability, we manually classified a stratified sample of 4,984 issues from a dataset of 12,910 issues across 36 Qiskit repositories. Automated classifications were compared with ground truth using accuracy, precision, recall, and F1-score. The framework achieved up to 85.21\% accuracy, with F1-scores ranging from 0.7075 (severity) to 0.8393 (quality attribute). Statistical validation via paired t-tests and Cohen's Kappa showed substantial to almost perfect agreement for bug type (k = 0.696), category (k = 0.826), quality attribute (k = 0.818), and quantum-specific bug type (k = 0.712). Severity classification showed slight agreement (k = 0.162), suggesting room for improvement. Large-scale analysis revealed that classical bugs dominate (67.2\%), with quantum-specific bugs at 27.3\%. Frequent bug categories included compatibility, functional, and quantum-specific defects, while usability, maintainability, and interoperability were the most impacted quality attributes. Most issues (93.7\%) were low severity; only 4.3\% were critical. A detailed review of 1,550 quantum-specific bugs showed that over half involved quantum circuit-level problems, followed by gate errors and hardware-related issues.
\end{abstract}


\section{Introduction}\label{sec1}

Quantum software engineering presents unique challenges that stem from the hybrid nature of quantum programs, which combine classical orchestration with quantum circuit execution. These challenges include managing the complexity of quantum operations, debugging under hardware noise, and dealing with limited observability in quantum states \cite{de2022software,gay2005communicating}. As the quantum software ecosystem matures, understanding and managing software defects becomes essential to ensure reliability, maintainability, and performance in quantum applications.
Automated bug classification plays a vital role in software quality assurance by enabling systematic defect analysis, supporting large-scale empirical studies, and informing tooling and development practices \cite{ yang2022comprehensive,jonsson2016automated}. In classical software domains, both rule-based and machine learning-based classifiers have been employed to categorize issues by type, severity, and quality impact with considerable success \cite{yang2022survey,ahmed2021capbug,bhandari2023buggin}. However, there is a significant gap in tools and techniques specifically designed to classify bugs in quantum software, where the nature and semantics of defects often differ markedly from classical counterparts\cite{ haghparast2023quantum,li2021understanding,de2022software,zhao2020quantum}.
In this study, we introduce a rule-based automated classification framework tailored to the quantum software domain. The framework classifies issues across five dimensions:
\begin{enumerate}
    \item 	Bug type (quantum or classical).
    \item 	Bug category (e.g., logical, syntax, quantum-specific).
    \item Severity (critical, high, medium, low).
    \item Impacted quality attribute (e.g., maintainability, reliability, performance).
    \item Type of quantum-specific issue (e.g., quantum gate errors, measurement errors, transpilation issues).
\end{enumerate}

This system uses curated keyword lists and heuristic rules grounded in domain knowledge of quantum computing and common failure modes in quantum programs. We collected 12,910 issues from the 36 repositories of Qiskit and performed the classification. To evaluate its effectiveness, we conducted a large-scale manual annotation of 4,984 GitHub issues from 36 repositories within the Qiskit ecosystem. Issues were selected using random stratified sampling to ensure representation across severity levels and bug types within this specific framework, but not across different categories of quantum software. Each issue was manually labeled for bug type, category, severity, and impacted quality attribute, forming a ground truth dataset for benchmarking.
We assessed the framework’s performance against this ground truth using standard metrics including accuracy, precision, recall, and F1-score, as well as inter-rater agreement via Cohen’s Kappa. We also employed paired t-tests to compare manual and automated classifications statistically.
Our findings demonstrate that the framework achieves substantial to near-perfect agreement ($\kappa > 0.69$) for bug type, category, and quality attribute classification. However, severity classification showed only slight agreement ($\kappa = 0.162$), suggesting it is more difficult to infer severity heuristically. These results highlight the framework’s utility for large-scale issue analysis in quantum software while also identifying areas for future enhancement.

This work offers a reproducible and interpretable approach to quantum bug classification and supports broader efforts to build robust quantum software analysis tools.
\subsection{Contributions}
This work makes the following key contributions toward improving automated bug classification in quantum software:
\begin{enumerate}
   \item \textbf{ A Rule-Based Framework for Multi-Dimensional Bug Classification:}
We propose a rule-based classification framework that automatically labels quantum software issues across five dimensions: bug type, bug category, severity level, impacted quality attribute, and quantum-specific subtype. The framework is designed to be interpretable, adaptable, and grounded in domain-specific keyword analysis.
\item\textbf{ A Manually Annotated Dataset of Quantum Software Issues:}
We construct a manually labeled dataset comprising 4984 issues from 36 Qiskit repositories, annotated along all five classification axes. This dataset serves as a benchmark for evaluating bug classification techniques in the quantum domain.
\item \textbf{Empirical Evaluation of Rule-Based Classification:}
We evaluate the classification performance of our rule-based system using standard metrics (precision, recall, F1-score) against the annotated dataset. Additionally, we apply a paired t-test to assess the statistical significance of performance differences and use Cohen's kappa to measure inter-rater agreement between the rule-based predictions and the human annotations. The results demonstrate that rule-based approaches can deliver high accuracy, strong agreement with human annotations, and interpretability in domains with limited labeled data and strong domain-specific characteristics.

\end{enumerate}
\subsection{Overall structure of the paper}
The paper is structured to provide a comprehensive exploration of a rule-based framework for automated classification of quantum software issues. Section 1 \ref{sec1} introduces the motivation and scope of the study, followed by a summary of the key contributions of the work. Section \ref{related work} reviews the related work, positioning the paper within the broader context of research on software bug classification and quantum software engineering. Section \ref{methodology}
presents the methodology, beginning with the process of dataset construction and manual annotation. It then outlines the multiple classification dimensions considered in this study, including classical vs. quantum classification, mapping of issues to bug categories, quality attributes, severity levels, and quantum-specific bug categories, while emphasizing transparency and reproducibility of the process.

Section \ref{results} reports the results of the evaluation. It includes a comparison of manual and automated classifications through statistical agreement, and presents classification performance metrics. The section also explores the distribution of bug types, bug categories, quality attributes, severity levels, and quantum-specific bug categories, along with a summary of the key findings. Section \ref{conclusion} offers the conclusion, summarizing the primary insights and outcomes of the study. Section \ref{future work} discusses directions for future work to extend and refine the framework. Finally, Section \ref{threats} addresses threats to validity, covering internal, external, and construct validity to assess the robustness and generalizability of the study's findings.

\section{Related Work}\label{related work}
Bug classification has been a longstanding area of interest in software engineering, with approaches ranging from manual taxonomies to automated machine learning models. Early work focused on categorizing bugs based on source code analysis, defect patterns, and fault types \cite{pan2009toward,kim2008classifying}. More recent studies have employed natural language processing (NLP) techniques to classify bugs using issue descriptions and commit messages \cite{dos2020commit, zhou2017automated, da2017using}. Tools like Bugzilla\cite{serrano2005bugzilla} and JIRA\cite{ortu2015jira} have served as standard platforms for collecting and analyzing bug reports across open-source ecosystems.
Rule-based systems have also been used in classical software for defect classification, particularly in industrial settings where interpretability and precision are crucial \cite{ahmad2021rule,wang2003fault}. These approaches often rely on keyword matching, regular expressions, or domain-specific ontologies to assign labels, allowing for easy customization and transparency in the classification logic.

\par
Quantum software is a nascent but rapidly evolving field, with unique challenges in debugging and reliability due to the probabilistic nature of quantum computation, hardware limitations, and unfamiliar programming abstractions \cite{heim2020quantum,ramalho2024testing}. A limited but growing body of research has begun to examine software quality issues in quantum programming frameworks such as Qiskit\cite{javadi2024quantum}, Cirq\cite{isakov2021simulations}, and PyQuil\cite{computing2019pyquil}.
Several studies have characterized quantum software bugs through manual inspection of GitHub issues and pull requests. For example, Palsodkar et al.(2023) \cite{palsodkar2023empirical} conducted an empirical analysis of Qiskit repositories and identified recurrent bug patterns, such as incorrect circuit construction or backend compatibility issues. More recently, research has also explored quantum-specific testing techniques\cite{ wang2021qdiff,pontolillo2025qucheck}, static analysis tools \cite{ zhao2023qchecker,assolini2024static}, and the role of simulators in mitigating defects \cite{ vovrosh2021simple,gely2020qucat}.
However, much of this prior work focuses on identifying or categorizing bugs at a coarse level. Few studies have attempted to build and evaluate automated systems for bug classification tailored to quantum software, particularly those that are interpretable and easy to validate by practitioners

Automated issue classification is a well-studied task in software repository mining. Techniques range from supervised learning using labeled datasets \cite { bissyande2013got}, to unsupervised topic modelling \cite{ sridhar2015unsupervised}, to hybrid approaches that combine rules with statistical models\cite{ villena2011hybrid, pourpanah2016hybrid}. For classical software, issue classification tasks typically include determining issue type (e.g., bug, enhancement, question), severity, and impacted quality attribute (e.g., performance, reliability).
While machine learning-based classifiers offer high accuracy, they often lack interpretability and require large volumes of labeled data, something not readily available in quantum software projects. Rule-based approaches, in contrast, offer greater transparency and adaptability, especially in emerging domains where domain knowledge is more valuable than raw data scale.

Our work contributes to this landscape by presenting a rule-based framework for multi-label classification of quantum software issues. Unlike prior efforts, our system performs fine-grained classification across multiple dimensions: bug type, category, severity, quality attribute, and quantum-specific subtype. We evaluate this framework on a manually annotated dataset of 4984 issues from 36 Qiskit repositories, demonstrating that rule-based systems can achieve reliable performance while remaining interpretable and adaptable to domain-specific needs.
This study fills a critical gap by providing the first comprehensive evaluation of automated bug classification tailored to the quantum domain, and offers insights into how rule-based techniques can support reliability engineering in quantum software development.

\section{Methodology}\label{methodology}
This study presents and evaluates a rule-based classification framework designed to automatically categorize software issues in quantum computing projects.Fig. \ref{fig:overallmethodology} provides an overview of the entire pipeline, while Fig. \ref{fig:rulebasedclassification} details the rule-based classification framework used for automated labeling. The framework addresses five key classification tasks: determining the bug type (quantum, classical, or uncategorized), categorizing the bug (e.g., logical, syntax, performance), assessing severity, identifying impacted quality attributes, and further sub-classifying quantum-specific issues. The working of each classification algorithm is given in subsection  \ref{classificationdim} . To assess the accuracy and reliability of the system, we manually annotated a benchmark dataset of 4,984 GitHub issues from 36 repositories within the Qiskit ecosystem. This annotated dataset serves as the ground truth for empirical evaluation of the automated system using standard classification metrics and statistical agreement measures.
\par

\subsection{Dataset Construction and Manual Annotation}
The dataset was constructed by extracting issues from 36 repositories belonging to the Qiskit project. These repositories encompass a variety of components within the Qiskit framework, including quantum circuit design libraries, simulators, backends, transpilers, and visualization tools. The selection of repositories ensures thematic consistency in the software domain while covering diverse functional aspects of a full-stack quantum computing library.

Using the GitHub REST API, we retrieved issue data including the issue number, GitHub ID, title, URL, state (open or closed), user-assigned labels, creation and closing timestamps, number of comments, and the full body of the issue description. Issues lacking descriptive content, containing duplicate information, or deemed irrelevant (e.g., pure feature requests or documentation typos without functional implications) were excluded. Textual data was preprocessed to remove markdown syntax, normalize casing and punctuation, and eliminate trivial formatting artifacts to facilitate downstream classification.

In total, we collected 12,910 issues from across the selected repositories. These were fed into our rule-based classification framework, which automatically labeled each issue across five dimensions: bug type, bug category, severity, impacted quality attribute, and quantum-specific subtype (for quantum-related bugs). To support rigorous evaluation, we manually annotated a stratified random sample of 4,984 issues. Stratification ensured proportional representation of repositories, bug types, and categories to avoid sampling bias.

Each issue in the sample was manually labeled across the five classification axes. First, we identified the bug type (quantum, classical, or uncategorized) based on whether the issue described faults in quantum-specific infrastructure (e.g., circuit transpilation, quantum execution backends) or conventional software components (e.g., CLI tools, data formatting). Second, issues were categorized into bug classes, such as logical errors, syntax issues, performance faults, or documentation deficiencies. These categories were informed by both established software defect taxonomies and the domain-specific nature of quantum programming. Third, severity levels were assigned based on inferred impact, urgency, and context, ranging from critical to low. Fourth, each issue was mapped to one or more quality attributes from the ISO/IEC 25010 standard \cite{estdale2018applying} and also as per the study \cite{sodhi2021quantum}, including maintainability, reliability, usability, performance efficiency, and interoperability. Finally, quantum-specific bugs were further annotated into fine-grained subtypes such as circuit construction errors, transpilation failures, backend misconfigurations, and simulator faults.

To assess the reliability of the classification of quantum-specific bugs, we additionally selected 30\% of all issues identified as quantum-related by the automated system and subjected this subset to manual review. This allowed for a focused evaluation of the system's capability to discern nuanced categories within quantum defect reports.

\begin{figure}[htbp]
    \centering
    \includegraphics[width=0.55\linewidth]{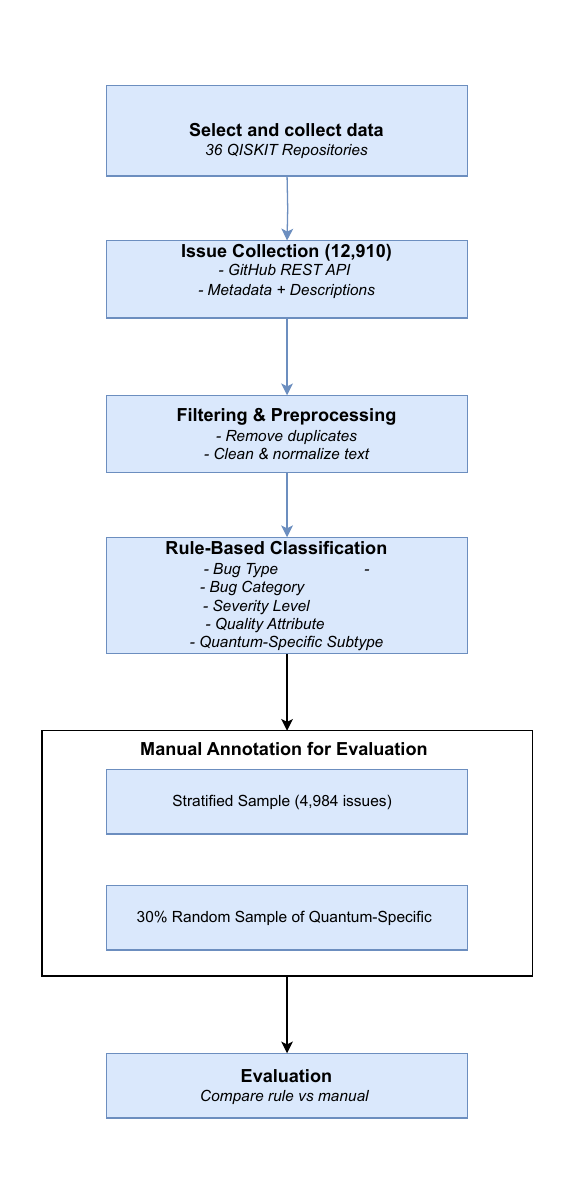}
    \caption{Overview of the data collection, classification, and evaluation pipeline. Issues were collected from 36 Qiskit repositories, filtered for relevance, automatically labeled across five dimensions, and evaluated using stratified and targeted manual annotation subsets.}
    \label{fig:overallmethodology}
\end{figure}

The full data processing and evaluation workflow is summarized in Figure ~\ref{fig:overallmethodology}. It shows the end-to-end pipeline from issue collection and filtering to rule-based classification, manual annotation, and evaluation.
The manual annotation was conducted by a team of researchers with expertise in quantum software development.
\subsection{Classification Dimensions} \label{classificationdim}
\begin{figure}[htbp]
    \centering
    \includegraphics[width=0.8\linewidth]{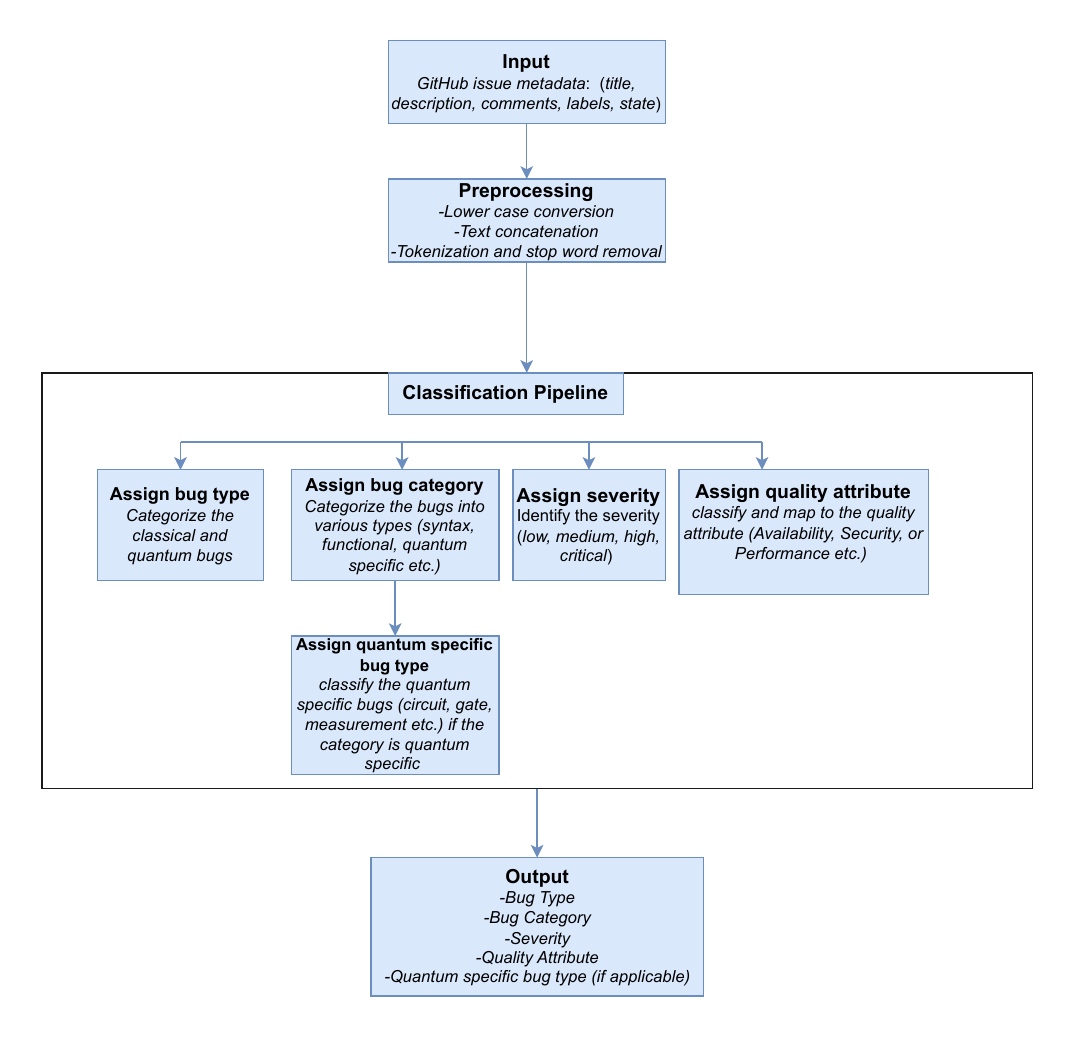}
    \caption{Rule based classification framework}
    \label{fig:rulebasedclassification}
\end{figure}

To gain deeper insights into the nature and impact of bugs in quantum software, we developed a multi-dimensional classification framework as shown in figure \ref{fig:rulebasedclassification}. This framework categorizes reported issues across several orthogonal dimensions, enabling a structured analysis of defects. Specifically, we classified bugs based on their category (e.g., functional, performance, compatibility), type (quantum-specific or classical), severity level, and the quality attribute they impact (such as maintainability or usability). Additionally, for quantum-specific bugs, we introduced a dedicated sub-categorization to capture domain-specific issues like circuit-level errors and transpilation problems. These dimensions together provide a comprehensive view of the defects affecting quantum software systems and form the basis for our subsequent empirical evaluation.

\subsubsection{Classical vs. Quantum Classification}

A "quantum bug" refers to unexpected errors or misbehavior in quantum computing systems, particularly those that are not easily detectable or fixable using classical error correction techniques. These bugs can arise from various sources, including software glitches, hardware imperfections, and even fundamental quantum phenomena \cite{campos2021qbugs}\cite{paltenghi2022bugs}. These bugs typically involve improper manipulation of qubits, incorrect gate sequences, or misapplication of quantum concepts like superposition, entanglement, and measurement.Consider the real-world issue shown in listing \ref{quantumbugex} from the Qiskit framework\cite{qiskitbug13732}.

\begin{lstlisting}[language=Python, caption={KeyError example in Qiskit's transpiler},label={quantumbugex}]
from qiskit import QuantumCircuit
from qiskit.compiler import transpile
qc = QuantumCircuit(2)
qc.h(0)
qc.cx(0, 1)

coupling_map = [[0, 1]]  # Only a single connection
transpiled_qc = transpile(qc, coupling_map=coupling_map)
final_layout = transpiled_qc.final_index_layout(filter_ancillas=False)
\end{lstlisting}
In this case, a \texttt{KeyError} is raised when invoking \texttt{final\_index\_layout(filter\_ancillas=False)} on certain transpiled quantum circuits. The error occurs due to the presence of disjoint coupling maps, where isolated qubits are not correctly tracked by the transpiler layout. This leads to issues during transpilation that prevent proper mapping of qubits to the hardware.

The error arises from a quantum-specific issue in handling coupling maps and qubit layouts, which directly affects the proper functioning of the quantum circuit transpilation process. Therefore, it is classified as a quantum bug.

A \textit{classical bug} in a quantum software system refers to an error that arises in the classical (non-quantum) parts of the program, such as data handling, control flow, or configuration logic. These bugs do not involve the misuse of quantum operations or qubit-level logic. Consider the real-world issue shown in listing \ref{classicbugex} from the Qiskit framework \cite{qiskitbug13730}.


In this case, the \texttt{StatevectorSampler} class incorrectly reinitializes the pseudorandom number generator with the same seed for each Program Unit Block (PUB). As a result, the outcomes of simulated quantum circuit executions are correlated, which violates the statistical expectations of independent runs.

\begin{lstlisting}[language=Python, caption={Seed reuse bug in Qiskit's StatevectorSampler},label={classicbugex}]
from qiskit import QuantumCircuit
from qiskit.primitives import StatevectorSampler
qc = QuantumCircuit(1)
qc.h(0)
qc.measure_all()
sampler = StatevectorSampler(seed=42)
res = sampler.run([(qc, None, 1) for _ in range(10)]).result()
for r in res:
    print(r.data.meas.get_int_counts())
\end{lstlisting}

 The issue arises from incorrect handling of a classical software component (random seed reuse), not from quantum-specific logic such as gate sequences or qubit manipulation. Therefore, it is classified as a classical bug.\par


To distinguish between quantum-specific and classical software issues, we developed a rule-based classification algorithm that analyzes the textual content of GitHub issues. Each issue is categorized as Quantum,\textbf{ \textbf{Classical}}, or Uncategorized based on a weighted scoring system applied to domain-specific terms and phrases.

\begin{algorithm}
\caption{Bug Type Classification Algorithm}
\begin{algorithmic}[1]
\Function{classify\_bug}{title, description, comments}
    \State text $\gets$ normalize(title + " " + description + " " + comments)
    \State quantum\_score, classical\_score $\gets$ 0, 0
    \For{each keyword in QUANTUM\_KEYWORDS}
        \If{keyword $\in$ text}
            \State quantum\_score += 1
        \EndIf
    \EndFor
    \For{each keyword in CLASSICAL\_KEYWORDS}
        \If{keyword $\in$ text}
            \State classical\_score += 1
        \EndIf
    \EndFor
    \For{each phrase in QUANTUM\_BIGRAMS}
        \If{phrase $\in$ text}
            \State quantum\_score += 2
        \EndIf
    \EndFor
    \For{each phrase in CLASSICAL\_BIGRAMS}
        \If{phrase $\in$ text}
            \State classical\_score += 2
        \EndIf
    \EndFor
    \If{negation\_phrase\_detected(text)}
        \State \Return ``Uncategorized''
    \EndIf
    \If{quantum\_score $>$ classical\_score}
        \State \Return ``Quantum''
    \ElsIf{classical\_score $>$ quantum\_score}
        \State \Return ``Classical''
    \Else
        \State \Return ``Uncategorized''
    \EndIf
\EndFunction
\end{algorithmic}
\end{algorithm}

\subsubsection{Text Aggregation and Preprocessing}

The algorithm begins by aggregating all textual components of an issue, including the title, description, and associated comments. The combined text is then normalized, converted to lowercase, and stripped of punctuation to ensure consistency in keyword matching and phrase detection.

\subsubsection{Keyword and Bigram Scoring}

A predefined set of \textit{unigrams} (individual keywords) and \textit{bigrams} (two-word phrases) is used for classification. These are grouped into two categories: \textbf{quantum-related} and \textbf{classical-related} terms. Each keyword is assigned an integer weight (1, 2, or 3) that reflects its specificity or relevance to its domain. Bigrams contribute a fixed score of 2, as they typically encode more context-specific signals (e.g., ``quantum algorithm'', ``compiler error'').
The algorithm scans the normalized issue text and maintains two parallel scores: \texttt{quantum\_score} and \texttt{classical\_score}. Each matched term contributes its weight to the corresponding score. In addition to core keywords, certain domain indicators such as ``Qiskit'' or ``Quantum SDK'' boost the \texttt{quantum\_score}, while terms like ``memory leak'' or ``CPU fault'' increase the \texttt{classical\_score}.

\subsubsection{Negation Handling}

To reduce false positives, the algorithm incorporates basic negation detection. If phrases such as ``not a quantum bug'' or ``not classical'' are identified in the text, the issue is immediately assigned to the \textbf{Uncategorized} category, overriding the score-based decision.

\subsubsection{Classification Decision Rules}

After processing the full text, the algorithm determines the issue’s type using the following logic:

\begin{itemize}
    \item If \texttt{quantum\_score} $>$ \texttt{classical\_score}, classify the issue as \textbf{Quantum}.
    \item If \texttt{classical\_score} $>$ \texttt{quantum\_score}, classify the issue as \textbf{Classical}.
    \item If the scores are equal or too low to indicate a dominant category, classify the issue as \textbf{Uncategorized}.
    \item If a negation phrase is detected, classify the issue as \textbf{Uncategorized} regardless of the scores.
\end{itemize}

\subsubsection{Transparency and Reproducibility}

This rule-based approach leverages domain-specific vocabulary and contextual phrases to differentiate between quantum and classical software bugs. To ensure reproducibility and facilitate future enhancements, the complete list of keywords, bigrams, and associated weights is made available in the supplementary documentation.

\begin{figure}
    \centering
    \includegraphics[width=0.99\linewidth]{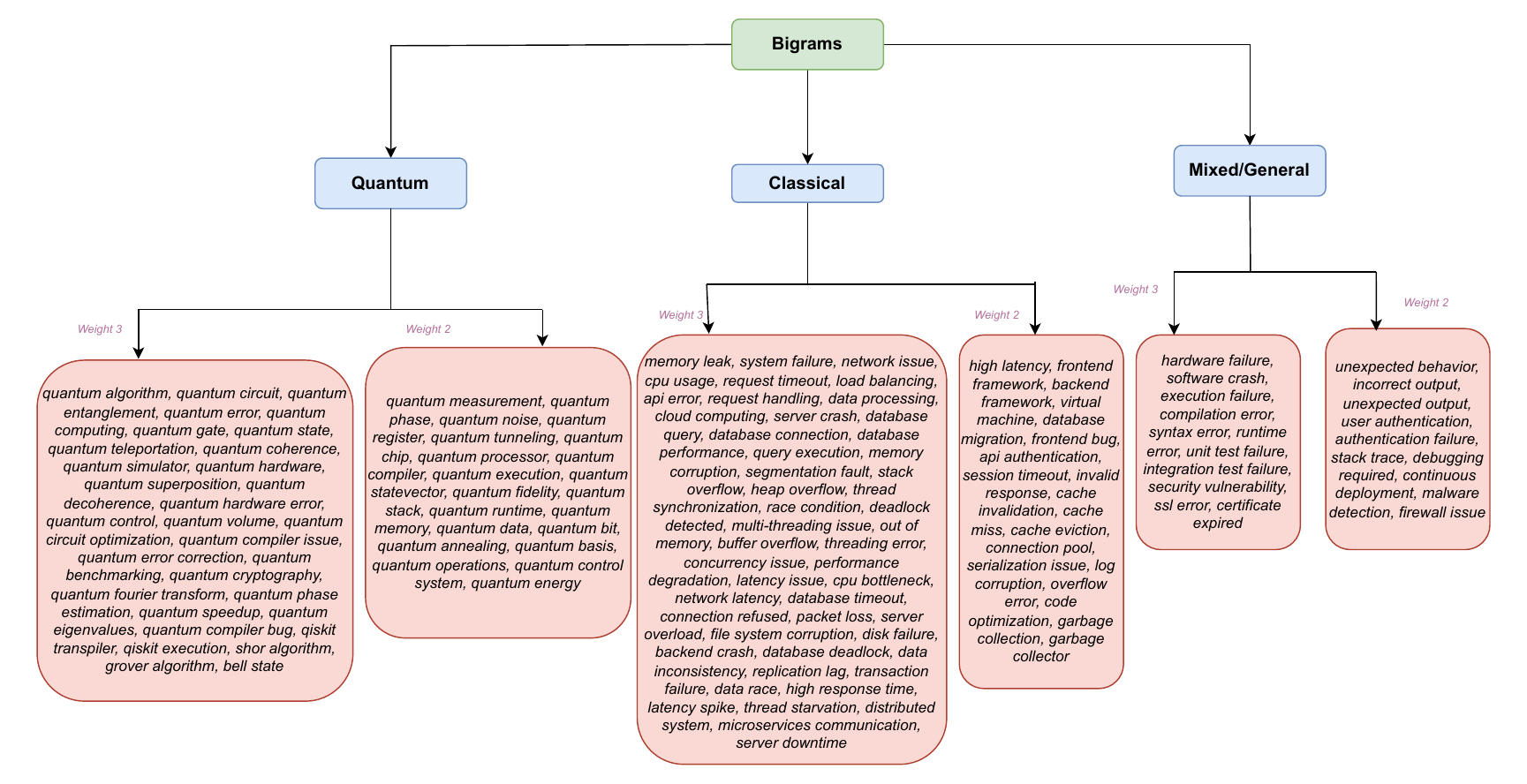}
    \caption{Categorization of bigram phrases used in bug classification. Phrases are classified into Quantum, Classical, and Mixed/General categories. Each phrase is assigned a weight (2 or 3) based on its semantic relevance. These bigrams provide context-rich cues that enhance classification accuracy.}
    \label{fig:bug_type_classification_bigrams}
\end{figure}

Figures~\ref{fig:bug_type_classification_unigrams} and~\ref{fig:bug_type_classification_bigrams} illustrate the unigram and bigram features used in the classification process, respectively. These figures group terms by domain and display their assigned weights, offering a comprehensive overview of the linguistic features underpinning the classification system.

\begin{figure}
    \centering
    \includegraphics[width=0.99\linewidth]{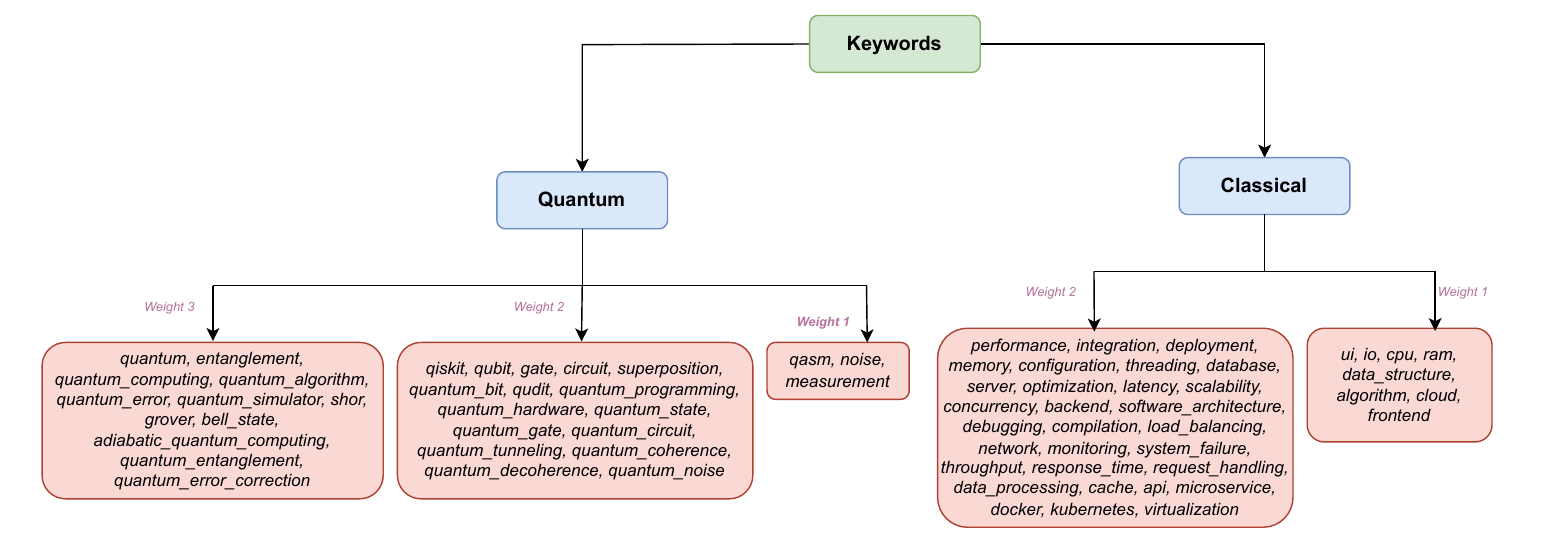}
    \caption{ Categorization of unigram keywords used for bug type classification. Keywords are grouped into Quantum and Classical domains and further organized by weight (1–3), which indicates their relative importance or specificity to their domain. Higher-weighted keywords contribute more to the classification score.}
    \label{fig:bug_type_classification_unigrams}
\end{figure}
\par
Each keyword and bigram used in the classification process is assigned a weight ranging from \textbf{1 to 3}, reflecting its \textit{domain specificity}, \textit{semantic strength}, and \textit{discriminative power}:

\begin{enumerate}
  \item Weight 3 is assigned to terms and phrases that are \textit{highly indicative} of a specific domain and \textit{rarely appear outside of it}. For example, terms like \texttt{quantum\_entanglement}, \texttt{shor algorithm}, or \texttt{memory leak} are strong signals that the issue is domain-specific, with minimal ambiguity.
  
  \item Weight 2 is used for terms that are \textit{moderately domain-specific}, frequently encountered in technical discussions within the domain, but may also appear in other contexts. Examples include \texttt{qubit}, \texttt{quantum\_state}, \texttt{database}, or \texttt{threading}.
  
  \item \textbf{Weight 1 }is reserved for \textit{general-purpose or common terms} that provide weaker evidence of domain relevance. These are often foundational terms (e.g., \texttt{measurement}, \texttt{ui}, \texttt{io}) that, on their own, may not reliably determine the nature of the bug.
\end{enumerate}

For bigrams, the weighting is generally higher (minimum 2), as \textit{multi-word phrases tend to carry more context} and therefore serve as stronger classification signals. For instance, \texttt{quantum error correction} or \texttt{network latency} provide much richer semantic cues than individual words like \texttt{error} or \texttt{latency}.
\subsubsection{ Bug Categories }
To enhance the granularity of issue analysis, we employ a structured bug categorization scheme that maps specific keywords and phrases to predefined bug types. This taxonomy covers a wide range of error classes commonly encountered in both quantum and classical computing contexts. Each bug report is evaluated for the presence of these diagnostic phrases, allowing the system to infer the most likely bug category based on semantic content.

\begin{figure}[h!]
    \centering
    \includegraphics[width=0.99\linewidth]{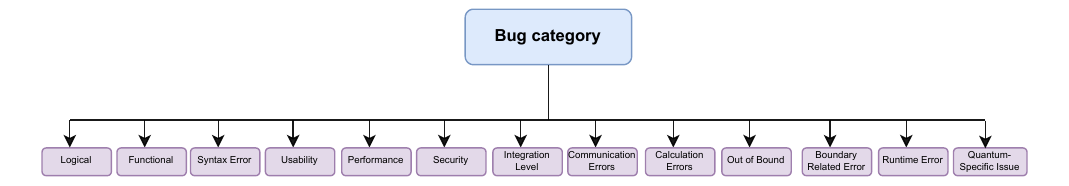}
    \caption{Categories of bugs}
    \label{fig:category_bug}
\end{figure}

The categorization includes 16 distinct bug types such as Logical, Functional, Syntax Errors, Performance, Security, and Quantum-Specific Issues, among others. These categories are used either independently or in conjunction with quality attribute and severity classification tasks. Each category is associated with a curated set of domain-specific keywords and indicative phrases (e.g., “memory leak,” “syntax error,” “quantum state collapse”). Figure \ref{fig:category_bug} shows the various categories of bugs.

We propose the Bug Categorization algorithm, which is designed to assign software issues to predefined categories using a combination of label inspection and keyword-based textual analysis. The process begins by collecting relevant textual components from each issue report, including the title, description, and user comments. If labels are attached to the issue, they are immediately checked against a predefined category mapping table. A matching label allows for immediate categorization, ensuring rapid classification where metadata is available.
\\
In the absence of relevant labels, the algorithm proceeds with a content-based analysis. All textual inputs are normalized by converting to lowercase, removing common stopwords, and applying lemmatization to unify variations of the same word. The normalized text is then evaluated against a dictionary of keywords associated with each predefined category. A score is computed for each category based on the frequency of keyword matches found in the issue’s text. The category with the highest score is selected and assigned to the bug. If no dominant category emerges from this scoring process, the issue is marked as "Uncategorized" to reflect uncertainty or ambiguity.
The Bug Categorization component is responsible for assigning software issues to one of several predefined categories. It operates by analyzing textual elements such as the title, description, and comments, and optionally using associated labels to enhance classification reliability. The process involves five primary steps:
\begin{enumerate}
    \item Data Collection: The algorithm extracts labels, the issue title, a detailed description, and any user comments for contextual understanding.
    \item Text Preprocessing: The collected text is normalized by converting it to lowercase, removing stopwords, and applying lemmatization to standardize word forms.
    \item Label-Based Category Assignment: If any of the issue labels match entries in a predefined category mapping, the corresponding category is assigned directly.
    \item Keyword-Based Scoring: In the absence of label matches, the algorithm computes a score for each category by checking for the presence of category-specific keywords in the normalized text.
    \item Final Category Decision: The category with the highest score is assigned to the issue. If no meaningful score is found, the issue is marked as “Uncategorized.”

\end{enumerate}
This method ensures that bugs are systematically categorized, even in cases where labels are incomplete or missing, thereby supporting more comprehensive issue tracking and analysis.

\begin{algorithm}[htbp] \label{algo:bug_categ}
\caption{Bug Categorization Algorithm}
\begin{algorithmic}[1]
\Function{assign\_category}{labels, title, description, comments}
    \State text $\gets$ normalize(title + " " + description + " " + comments)
    \For{each label in labels}
        \If{label $\in$ CATEGORY\_MAPPING}
            \State \Return CATEGORY\_MAPPING[label]
        \EndIf
    \EndFor
    \State category\_scores $\gets$ initialize\_scores()
    \For{each category in CATEGORY\_KEYWORDS}
        \For{each keyword in CATEGORY\_KEYWORDS[category]}
            \If{keyword $\in$ text}
                \State category\_scores[category] += 1
            \EndIf
        \EndFor
    \EndFor
    \State max\_category $\gets$ category with highest score
    \If{max\_category has significant score}
        \State \Return max\_category
    \Else
        \State \Return ``Uncategorized''
    \EndIf
\EndFunction
\end{algorithmic}
\end{algorithm}
\subsubsection{Quality Attribute Mapping}
To identify the category of the quality attributes, we developed an algorithm in which the \texttt{assign\_quality\_attributes} function is developed to classify an issue into one of several predefined quality attributes, such as \textit{Availability}, \textit{Security}, \textit{Performance} and others \cite{sodhi2021quantum}. This classification is performed by analyzing the textual content of the issue, including its title, body, and comments. To determine the most relevant quality attribute, the function uses a combination of techniques: keyword matching, TF-IDF (Term Frequency--Inverse Document Frequency) scoring, and weighted keyword frequency analysis. Together, these methods provide a layered approach to accurately categorizing issues based on the language used within them.
\begin{algorithm}
\caption{assign\_quality\_attributes}
\begin{algorithmic}[1]
\State Initialize \texttt{QUALITY\_ATTRIBUTE\_MAPPING} with quality attributes and associated keywords
\State Initialize \texttt{QUALITY\_PRIORITY} list with the order of quality attributes
\State Extract \texttt{labels} from the issue (if any) and convert them to lowercase
\State Combine the issue's title, body, and comments into a single string \texttt{combined\_text}
\State Convert \texttt{combined\_text} to lowercase

\Function{preprocess\_text}{text}
    \State Tokenize the \texttt{text}
    \State Remove stopwords
    \State \Return the filtered \texttt{text}
\EndFunction

\Function{match\_keywords}{text, keywords}
    \State Preprocess the \texttt{text} by removing non-alphanumeric characters
    \State Tokenize the preprocessed \texttt{text}
    \State Create a word count of the \texttt{text}
    \For{each \texttt{keyword} in \texttt{keywords}}
        \If{\texttt{keyword} exists in the word count}
            \State \Return \texttt{True}
        \EndIf
    \EndFor
    \State \Return \texttt{False}
\EndFunction
\Function{tfidf\_keywords}{text, keywords}
    \State Use TF-IDF vectorizer with the given \texttt{keywords}
    \State Create a TF-IDF matrix for the \texttt{text}
    \State Extract non-zero TF-IDF scores
    \State \Return the sum of the TF-IDF scores
\EndFunction

\Function{weighted\_match}{text, keywords}
    \State Initialize \texttt{weight\_sum} to 0
    \For{each \texttt{keyword} in \texttt{keywords}}
        \State Count occurrences of the \texttt{keyword} in the \texttt{text}
        \State Multiply the count by the weight of the \texttt{keyword} (default to 1)
        \State Add the result to \texttt{weight\_sum}
    \EndFor
    \State \Return \texttt{weight\_sum}
\EndFunction

\For{each \texttt{quality} in \texttt{QUALITY\_PRIORITY}}
    \State Get the list of \texttt{keywords} for the current quality from \texttt{QUALITY\_ATTRIBUTE\_MAPPING}
    \State Create a dictionary with \texttt{keyword\_weights} (default weight = 1)
    \State Compute the \texttt{weighted\_sum} by applying \texttt{weighted\_match} on \texttt{combined\_text}
    \If{The issue’s labels match the quality keywords \textbf{or} the TF-IDF score is above threshold \textbf{or} \texttt{weighted\_sum} is greater than 0}
        \State \Return the \texttt{quality} as the assigned attribute
    \EndIf
\EndFor
\State \Return ``Miscellaneous''
\end{algorithmic}
\end{algorithm}
The function begins by initializing the \texttt{QUALITY\_ATTRIBUTE\_MAPPING} dictionary, which links each predefined quality attribute, such as Availability, Security, or Performance, to a curated list of associated keywords. These keywords represent common terms that signal the presence of specific quality concerns in issue descriptions. For example, “downtime” and “uptime” relate to Availability, while “encryption” and “authentication” are tied to Security. This mapping enables the algorithm to recognize and classify issues based on textual cues that reflect various software quality attributes.\par
The second structure is the \texttt{QUALITY\_PRIORITY} list, which specifies the order in which attributes should be evaluated. This ordered list is essential for prioritizing certain attributes over others when multiple conditions are satisfied.
Next, the function extracts useful information from the GitHub issue. It retrieves the issue's labels, which often serve as hints to its underlying quality dimension (e.g., a label like ``usability'' suggests a concern with user interface or experience). It also concatenates the issue’s title, body, and comments into a single string referred to as \texttt{combined\_text}. This aggregation helps capture the full context of the issue, as critical keywords might appear across different parts of the conversation. \par
Once the text is extracted, it undergoes preprocessing to prepare it for analysis. This involves tokenization, where the text is split into individual words, and stopword removal, which filters out common words like ``the'' or ``is'' that do not contribute meaningful information. The result is a cleaner, more informative version of the text that highlights key terms and improves the effectiveness of keyword matching and scoring techniques.
The function then checks whether any of the predefined keywords associated with each quality attribute are present in the cleaned text. This is achieved through a helper function called \texttt{match\_keywords}, which tokenizes the text and looks for overlaps with the attribute’s keyword list. A match at this stage suggests potential relevance but is not yet conclusive.
To further refine the classification, the function employs the TF-IDF technique. By analyzing the frequency of terms within the issue relative to a broader corpus of issues, TF-IDF scoring identifies keywords that are contextually important. A keyword that appears multiple times within a specific issue but is uncommon in the rest of the dataset will receive a high TF-IDF score. If any keyword associated with a quality attribute surpasses a threshold TF-IDF value (e.g., 0.1), it strongly indicates the relevance of that attribute.

In addition to simple keyword presence and TF-IDF scoring, the function also uses a weighted keyword matching mechanism. Here, each keyword is assigned a weight (typically defaulting to 1 unless specified otherwise), and the number of times it appears in the issue text is multiplied by this weight. The results are summed to produce a total weighted score. This score helps emphasize attributes whose keywords occur more frequently and with greater weight, further supporting classification.

With all these components in place, the function proceeds to assign a quality attribute. It loops through the attributes in the order defined by \texttt{QUALITY\_PRIORITY} and checks for a match based on three criteria: whether any labels match the attribute’s keywords, whether the TF-IDF score for the attribute's keywords exceeds the threshold, and whether the weighted keyword sum is greater than zero. If any of these checks pass, the corresponding attribute is assigned to the issue. If none of the attributes meet the conditions, the issue is categorized as ``Miscellaneous,'' indicating that it doesn’t clearly map to any specific quality attribute.

To demonstrate the behavior of the function, consider various issue scenarios. An issue mentioning “system downtime” and labeled “availability” would be categorized under Availability, while reports referencing vulnerabilities or exploits would map to Security. Complaints about “slow execution” or “latency” are classified as Performance, and issues involving “incompatibility” with external tools fall under Interoperability. Mentions of missing tests or debugging challenges point to Testability, whereas crashes and instability indicate Reliability. Feature requests related to monitoring or configuration suggest Manageability, and scalability concerns, such as high-load failures, are labeled Scalability. Comments on poor documentation or the need for refactoring reflect Maintainability. When no specific quality dimension is evident, the issue is categorized as Miscellaneous.

This algorithm provides a robust and intelligent method for classifying issues by analyzing natural language content. By combining keyword presence, contextual weighting, and prioritization logic, it maps issues to quality attributes such as \textit{Availability}, \textit{Security}, \textit{Performance}, \textit{Usability}, \textit{Maintainability}, or \textit{Testability}, facilitating better organization, analysis, and prioritization of software issues.

\subsubsection{Severity Identification}

To identify the severity of the bugs, we propose an algorithm in which the assign\_severity function is designed to determine the severity level of an issue in a repository. It does this by analyzing several components of the issue, including its labels, textual content (title, body, and comments), the number of comments, and the issue’s state (e.g., open or closed). This multilayered approach allows for a more context-aware severity classification, going beyond simple keyword detection to incorporate metadata and heuristic rules. Below is a detailed explanation of each step in the process, along with illustrative examples and pseudocode.

The process begins by extracting labels from the issue. If labels are present, they are collected and normalized to lowercase. This normalization ensures consistent comparison against a predefined dictionary of severity labels. For instance, if an issue has the labels ["Bug", "Critical"], they are converted to ["bug", "critical"]. These labels are later used to directly map to corresponding severity levels such as "Low", "High", or "Critical".

Next, the function constructs a unified text string from the issue’s title, body, and its comments. This aggregated text is transformed to lowercase to allow for case-insensitive keyword matching. For example, consider an issue with the title “Error in Quantum Algorithm,” a body stating “The algorithm fails when running on certain inputs,” and comments such as “I think this might be a bug.” All these components are concatenated and processed into a single lowercase string. This combined text is then preprocessed, typically involving tokenization and stopword removal to prepare it for keyword analysis.

With the text prepared, the function initializes the severity to “Low” as a default fallback value. If no subsequent conditions are satisfied, this will be the final assigned severity.

Following initialization, the function checks whether any of the labels match entries in a predefined dictionary called SEVERITY\_LABELS. For example, if SEVERITY\_LABELS maps "bug" to "Low" and "critical" to "Critical", the presence of the label "critical" will result in an immediate severity assignment of "Critical". This step allows developers to directly control severity through labeling conventions.

If no relevant severity is assigned via labels, the function proceeds to examine the issue text for specific severity-indicating keywords. These keywords are grouped by severity level and paired with weights to account for the strength or significance of the term. For example, the keyword “bug” might be weighted more heavily than “issue” for low severity, and “failure” might have a high weight for "High" severity. If a keyword is found in the text, its weight is added to a running total for its corresponding severity level. At the end of this process, the severity level with the highest cumulative score is selected.\par
To support keyword-based severity inference, the function uses a predefined dictionary of severity-specific keywords and their associated weights. These are grouped into categories (critical, high, medium and low ) each containing representative terms that commonly signal issues of that severity. For instance, the Critical category includes high-impact keywords like "crash", "system down", "data corruption", and "security breach", each assigned a weight of 4 or 5 to reflect their severity. Similarly, High severity includes terms such as "data loss", "major bug", and "stability", while Medium covers phrases like "performance", "lag", or "timeout". Finally, Low severity is associated with more cosmetic or non-functional terms such as "UI", "typo", or "visual". These weights guide the algorithm in scoring each category and determining the final severity classification based on the highest cumulative score. This weighting mechanism ensures that more serious keywords dominate the decision when multiple severities are mentioned.\par
In cases where the keyword-based score is still insufficient to assign a meaningful severity, the function evaluates the number of comments on the issue. If the issue has more than 10 comments and the current severity is still “Low”, the severity is elevated to “Medium”. If it exceeds 20 comments, and the current severity is either “Low” or “Medium”, it is raised to “High”. This heuristic assumes that the amount of discussion reflects issue complexity or importance.

Security-related considerations are also incorporated into the classification logic. If the label list includes "security" or the combined issue text contains the word "vulnerability", the severity is immediately set to "Critical", regardless of any previously assigned value. This ensures that potentially dangerous issues receive the highest attention.

Finally, the function checks the state of the issue. If it is marked as "closed", "resolved", or "fixed" and the severity is not already "Critical", then the severity is reset to "Low". This step reflects the assumption that resolved issues no longer require urgent attention, unless they were of critical nature.

The following walkthrough demonstrates the function’s end-to-end behavior. Consider an issue titled “Error in Quantum Algorithm,” with a body describing the algorithm’s failure on certain inputs, a label of “bug,” and 15 comments. The label "bug" maps to "Low", and keywords like “error” and “bug” further support this classification. However, due to the number of comments (15), the severity is elevated to “Medium”. If the issue also had a “security” label, the severity would be upgraded to “Critical”.
\begin{algorithm}[h!]
\caption{assign\_severity}
\begin{algorithmic}[1]
\State \textbf{Input:} \texttt{issue}, \texttt{comments}
\State \textbf{Output:} Severity of the issue

\State \texttt{labels} $\gets$ List of labels in \texttt{issue} converted to lowercase
\State \texttt{combined\_text} $\gets$ Concatenate \texttt{title}, \texttt{body}, and \texttt{comments} of \texttt{issue}, all in lowercase
\State \texttt{tokens} $\gets$ Preprocess \texttt{combined\_text} (e.g., tokenize, remove stopwords)

\State \texttt{severity} $\gets$ "Low"
\ForAll{label in \texttt{labels}} 
    \If{label is in \texttt{SEVERITY\_LABELS}} 
        \State \texttt{severity} $\gets$ \texttt{SEVERITY\_LABELS[label]}
        \State \textbf{break}
    \EndIf
\EndFor
\If{\texttt{severity} = "Low"} 
    \State \texttt{severity\_scores} $\gets$ Empty Counter
    \ForAll{(sev\_level, keywords) in \texttt{KEYWORDS}} 
        \ForAll{(keyword, weight) in \texttt{keywords}} 
            \If{keyword in \texttt{combined\_text}}
                \State \texttt{severity\_scores[sev\_level]} += weight
            \EndIf
        \EndFor
    \EndFor
    \If{\texttt{severity\_scores} is not empty}
        \State \texttt{severity} $\gets$ Severity with maximum score in \texttt{severity\_scores}
    \EndIf
\EndIf

\If{\texttt{issue.comments} $>$ 10 and \texttt{severity} = "Low"}
    \State \texttt{severity} $\gets$ "Medium"
\ElsIf{\texttt{issue.comments} $>$ 20 and \texttt{severity} in ["Low", "Medium"]}
    \State \texttt{severity} $\gets$ "High"
\EndIf

\If{"security" in \texttt{labels} or "vulnerability" in \texttt{combined\_text}}
    \State \texttt{severity} $\gets$ "Critical"
\EndIf

\State \texttt{state} $\gets$ \texttt{issue.state} (lowercase)
\If{\texttt{state} in ["closed", "resolved", "fixed"] and \texttt{severity} != "Critical"}
    \State \texttt{severity} $\gets$ "Low"
\EndIf

\State \textbf{Return} \texttt{severity}
\end{algorithmic}
\end{algorithm}

\subsubsection{Quantum-Specific Bug Categories}
Quantum computing introduces distinct challenges in software engineering due to its complex and probabilistic nature. Unlike classical systems, quantum programs interact with hardware and physics at a much deeper level, making debugging and issue categorization significantly more difficult. To streamline the process of analyzing quantum-specific issues reported in software repositories, we propose a \textbf{Quantum Bug Classification Algorithm} that automatically assigns a high-level category to quantum-related bug reports. Based on prior studies \cite{campos2021qbugs,zhao2021identifying,zhao2023bugs4q}, we classify quantum-specific bugs into the following categories:
\begin{enumerate}
    \item \textbf{Quantum Circuit Issues:} These involve incorrect circuit construction such as misconfigured qubit registers, incorrect initialization, or faulty state preparation. Bugs in this category may result in logically incorrect circuits that do not reflect the intended quantum operation.
    \item \textbf{Quantum Gate Errors:} Arising from the misuse of quantum gates like CNOT, Hadamard, or parameterized unitaries, these bugs often lead to unintended transformations or invalid quantum states. Zhao et al. \cite{zhao2021identifying} identified these as common in quantum algorithm implementations.
    \item \textbf{Quantum Measurement Errors:} This class includes bugs related to measurement basis errors, poor shot sampling, and misinterpretation of readout results, potentially leading to inaccurate outputs or invalid quantum state collapse \cite{zhao2023bugs4q}.
    \item \textbf{Quantum Noise Issues:} Due to the fragile nature of quantum states, bugs may arise from unaddressed noise, decoherence, or crosstalk between qubits. These are particularly problematic in NISQ-era devices, and often require mitigation strategies \cite{zhao2023bugs4q}.
    \item \textbf{Quantum Transpilation Issues:} These stem from improper transformation of circuits to conform to hardware topologies. Bugs include illegal qubit mappings, excessive SWAP gates, or compilation failures during transpilation \cite{campos2021qbugs}.
    \item \textbf{Quantum Algorithm Implementation Bugs:} High-level algorithm bugs, such as incorrect construction of Grover’s oracle or parameter misconfiguration in VQE, are prominent. These can break the logic of variational or amplitude amplification-based algorithms \cite{zhao2023bugs4q}.
    \item \textbf{Quantum Resource Constraints:} Bugs in this category include exceeding available qubits, memory overflow on simulators, or trying to run circuits too large for the target backend, which often causes runtime crashes \cite{zhao2023bugs4q}.
    \item \textbf{Quantum Entanglement and Decoherence Issues:} These are conceptual bugs where intended entanglement (e.g., Bell or GHZ states) is disrupted due to noise or gate misapplication, affecting fidelity or causing spurious correlations \cite{zhao2023bugs4q}.
    \item \textbf{Quantum Hardware-Specific Bugs:} Some bugs are specific to particular quantum devices or backends, such as IBM Q, Rigetti, or IonQ, due to backend configuration mismatches or device calibration errors \cite{campos2021qbugs}.
    \item \textbf{Quantum Error Correction (QEC) issues:} QEC-related bugs arise in surface code implementations, syndrome decoding, or stabilizer operations. While less frequent in NISQ applications, they are crucial for fault-tolerant systems \cite{zhao2021identifying}.
    \item \textbf{Hybrid Quantum-Classical Interface Issues:} Bugs here stem from the interface between classical optimizers and quantum circuits, particularly in hybrid algorithms like VQE or QAOA. These include data conversion errors, shape mismatches, or execution flow bugs \cite{zhao2023bugs4q}.
\end{enumerate}
The proposed algorithm combines \textit{keyword-based matching} and \textit{semantic similarity (TF-IDF)} to detect patterns in issue descriptions and map them to a set of curated quantum bug categories. This facilitates faster triaging, better routing to domain experts, and the development of more targeted tooling for quantum software quality assurance.


The following categories as shown in table \ref{tab:quantumbugcategorykeywords} represent common types of bugs observed in quantum software development. Each category is associated with representative keywords and a corresponding weight reflecting its relative importance:

\begin{table}[h!]
\centering
\caption{Quantum Bug Categories, Keywords, and Assigned Weights}
\begin{tabular}{|p{5cm}|p{7cm}|c|}
\hline
\textbf{Category} & \textbf{Keywords} & \textbf{Weight} \\
\hline
Quantum Circuit Issues & circuit, qubit, register, entanglement, superposition, state preparation, quantum register mapping, circuit optimization & 1.2 \\
\hline
Quantum Gate Errors & cnot, hadamard, t-gate, toffoli, swap, rz, rx, ry, uccsd, incorrect gate, wrong unitary, gate calibration & 1.3 \\
\hline
Quantum Measurement Errors & measurement, readout, basis, state collapse, shot noise, wrong measurement basis, incorrect probability distribution & 1.2 \\
\hline
Quantum Noise Issues & noise, decoherence, gate fidelity, error mitigation, quantum volume, fault tolerance, crosstalk, leakage error & 1.4 \\
\hline
Quantum Transpilation Issues & transpiler, swap, circuit mapping, qubit connectivity, gate translation, compilation, hardware constraints & 1.2 \\
\hline
Quantum Algorithm Implementation Bugs & grover, shor, qaoa, vqe, qft, amplitude amplification, phase estimation, oracle construction & 1.5 \\
\hline
Quantum Resource Constraints & qubit limit, out of memory, register overflow, resource error, circuit too large, hardware capacity exceeded & 1.1 \\
\hline
Quantum Entanglement and Decoherence Issues & entanglement, decoherence, state fidelity, bell state, ghz state, noise-induced state collapse, spurious correlations & 1.3 \\
\hline
Quantum Hardware-Specific Bugs & ibm quantum, rigetti, ionq, hardware error, backend issue, device calibration, chip connectivity, hardware fidelity & 1.2 \\
\hline
Quantum Error Correction (QEC) Bugs & error correction, syndrome decoding, logical qubit, surface code, stabilizer, quantum parity check & 1.4 \\
\hline
Hybrid Quantum-Classical Interface Issues & classical-quantum, hybrid, interface error, data conversion, tensor network, quantum kernel, variational circuit & 1.3 \\
\hline
\end{tabular}
\label{tab:quantumbugcategorykeywords}
\end{table}

The weights assigned to each category reflect their importance in the classification process. They are chosen based on the following factors:

\begin{enumerate}
    \item \textbf{Technical Criticality}: Some categories such as ``Quantum Algorithm Implementation Bugs'' and ``Quantum Error Correction Bugs'' directly affect the correctness of quantum computations, thus receiving higher weights.
    \item \textbf{Frequency in Bug Reports}: Categories with a higher frequency of occurrence in real-world issue reports, such as noise and circuit issues, are moderately weighted to reflect their relevance.
    \item \textbf{Debugging Complexity}: Categories that are harder to diagnose and fix, such as hardware-related or hybrid interface issues, are given slightly higher priority.
    \item \textbf{Impact on Fidelity and Reliability}: Issues affecting state fidelity, noise, and decoherence have higher consequences and thus are assigned larger weights.
\end{enumerate}
\par

The proposed algorithm  adopts a hybrid approach that leverages both keyword-based scoring and semantic similarity using the TF-IDF method. The process begins with a text preprocessing stage. In this stage, the input issue text is first converted to lowercase to ensure uniformity. Special characters and punctuation are removed to reduce noise in the text. Finally, stemming is applied using the Porter Stemmer to reduce each word to its root form, allowing for more robust keyword matching regardless of grammatical variation.

Following preprocessing, the algorithm performs keyword-based matching. For each predefined quantum bug category, it checks how many of the category’s stemmed keywords appear in the stemmed issue text. Each match is weighted by the category’s importance factor, resulting in a weighted keyword score for that category.

To enhance the robustness of the classification, the algorithm then incorporates semantic similarity through TF-IDF (Term Frequency–Inverse Document Frequency). It creates a text corpus where each document represents a category constructed by concatenating its associated keywords. The cleaned issue text is appended to this corpus. A TF-IDF vectorizer is then used to convert all documents into numerical vectors, which are compared using cosine similarity. The resulting similarity scores are scaled by a constant (e.g., 2.0) and added to the keyword scores.

In the final step, the algorithm sums the combined scores (keyword-based and TF-IDF-based) for each category. The category with the highest total score is selected as the best match. If all categories yield a score of zero, the issue is classified as “Unclassified,” indicating a lack of sufficient evidence for a meaningful categorization.
\begin{algorithm}\label{classifyquantumbugcategory}
\caption{Classify Quantum Bug Category}
\begin{algorithmic}[1]
\Require Issue text $T$
\Ensure Most relevant quantum bug category or \texttt{Unclassified}

\State Define a set of categories $C$, each with:
    \begin{itemize}
        \item a list of keywords $K_c$
        \item a weight $w_c$
    \end{itemize}
\State Apply stemming to all keywords $K_c$ for each category

\Function{CleanText}{$T$}
    \State Convert $T$ to lowercase
    \State Remove special characters (retain alphanumeric, spaces, dashes)
    \State Apply stemming to each word
    \State \Return cleaned and stemmed text $T'$
\EndFunction

\State $T' \gets \text{CleanText}(T)$
\State Initialize empty score map $S$

\ForAll{category $c$ in $C$}
    \State Count keyword matches $m_c$ between $T'$ and stemmed $K_c$
    \State $S[c] \gets m_c \times w_c$
\EndFor

\State Create a TF-IDF corpus with all category keyword sets and $T$
\State Compute TF-IDF vectors and cosine similarities between $T$ and each category $c$
\ForAll{category $c$ in $C$}
    \State $S[c] \gets S[c] + 2 \times \text{TFIDF\_similarity}(T, c)$
\EndFor

\If{all scores in $S$ are 0}
    \State \Return \texttt{Unclassified}
\Else
    \State \Return category $c^*$ where $S[c^*] = \max_{c \in C} S[c]$
\EndIf

\end{algorithmic}
\end{algorithm}

\section{Results}\label{results}

This section presents the results of evaluating the automated rule-based classification system against a manually annotated dataset. In total, 12,910 issues were collected from 36 Qiskit repositories and automatically classified across five dimensions: bug type, severity, category, impacted quality attribute, and quantum-specific bug subtype.

To assess the effectiveness of the automated system, a stratified random sample of 4,984 issues was selected from the full dataset and manually labeled. These manually annotated issues serve as the ground truth for evaluating classification performance.

Before delving into performance metrics, we first present a descriptive analysis of the entire automatically classified dataset. This analysis provides insight into the distribution of bugs across key dimensions, including classical vs. quantum categorization, bug categories (e.g., logic, API misuse), severity levels (e.g., high, medium, low), and impacted quality attributes (e.g., reliability, usability). For quantum-specific bugs, we also break down the distribution of finer-grained subtypes, such as circuit, simulator, and transpiler-related issues.

Following the descriptive overview, we evaluate the classification system's performance on the manually labeled subset. Performance is reported for each classification task using Cohen’s Kappa to assess agreement, paired t-tests to evaluate statistical significance, and standard classification metrics such as accuracy, precision, recall, and F1-score. Visualizations and summary tables are included to highlight key trends and findings.
\subsection{Evaluation}

To evaluate the effectiveness of the rule-based classification framework, a total of 4,984 GitHub issueswere manually annotated across five classification dimensions: \textit{bug type}, \textit{severity}, \textit{bug category}, \textit{quality attribute}, and \textit{quantum-specific bug type}. These issues were drawn from 36 Qiskit repositories, using stratified sampling to preserve distributional characteristics across repositories. A focused subset of 1,550 quantum-specific bugs was also curated to assess the system's ability to identify types of quantum-related issues. From this subset, a 30\% stratified sample was used for evaluating quantum-specific bug type classification.

The automated system's outputs were compared with manual annotations using \textit{paired t-tests} and \textit{Cohen’s Kappa} to assess statistical agreement. Additionally, standard classification metrics such as \textit{accuracy}, \textit{precision}, \textit{recall}, and \textit{F1-score} were calculated to evaluate performance against the ground truth.

\subsubsection{Statistical Agreement Between Manual and Automated Classifications}

A statistical comparison was conducted to determine how closely the automated classifications aligned with human annotations. \textit{Cohen’s Kappa} was used to measure agreement beyond chance, and \textit{paired t-tests} assessed the significance of differences across the five classification tasks.

As shown in Table~\ref{tab:ttest-kappa}, severity classification exhibited the weakest agreement with a Kappa value of 0.162, indicating only slight consistency between manual and automated labels. This was accompanied by a highly significant difference ($p < 0.0001$), suggesting systematic discrepancies in this attribute. In contrast, \textit{bug category} and \textit{quality attribute} classifications demonstrated almost perfect agreement, with Kappa scores of 0.826 and 0.818, respectively. Bug type and quantum-specific bug type classifications showed \textit{substantial agreement}, with Kappa values above 0.69, indicating that the automated system performs well in recognizing these dimensions.
\newcolumntype{P}[1]{>{\raggedright\arraybackslash}p{#1}}
\begin{table}[h!]
\centering
\caption{Paired t-Test and Cohen's Kappa for Classification Comparison}
\label{tab:ttest-kappa}
\begin{tabular}{|l|l|P{2cm}|P{4cm}|}
\hline
\textbf{Attribute} & \textbf{Paired t-Test} & \textbf{Cohen's Kappa} & \textbf{Interpretation} \\ 
\hline
Severity & $t = -30.0853$, $p < 0.0001$ & 0.162 & Slight agreement with significant deviation. Indicates need for refinement. \\ \hline
Bug Type & $t = 17.2721$, $p < 0.0001$ & 0.696 & Substantial agreement. Robust performance by the automated system. \\ \hline
Category & $t = -3.1290$, $p = 0.0018$ & 0.826 & Almost perfect agreement. High reliability. \\ \hline
Quality Attribute & $t = -1.9996$, $p = 0.0456$ & 0.818 & Almost perfect agreement with marginally significant difference. \\ \hline
Quantum-Specific Bug Type & $t = 9.4321$, $p < 0.0001$ & 0.712 & Substantial agreement. Good classification performance for quantum-specific bugs. \\ \hline
\end{tabular}
\end{table}

\subsubsection{Classification Performance Metrics}

The rule-based classifier was also evaluated using standard performance metrics. Table~\ref{tab:performance-metrics} reports the \textit{accuracy}, \textit{weighted precision}, \textit{weighted recall}, and \textit{weighted F1-score} for each classification dimension.

The system achieved its highest performance in \textit{quality attribute classification}, with an accuracy of 85.21\% and an F1-score of 83.93\%. \textit{Bug category} and \textit{bug type} classifications also performed strongly, with F1-scores exceeding 83\% and accuracy levels above 84\%. Classification of \textit{quantum-specific bug types} achieved an accuracy of 83.75\%, demonstrating that the rule-based system is effective in capturing quantum-specific domain knowledge.

On the other hand, \textit{severity classification} continued to be a relative weakness. Despite achieving a reasonable accuracy of 78.73\% , the F1-score was lower at 70.75\%, reflecting the system’s difficulty in consistently identifying severity levels, especially in cases where issue descriptions lacked explicit urgency cues.

\begin{table}[h!]
\centering
\caption{Classification Metrics Compared to Manual Labels}
\label{tab:performance-metrics}
\begin{tabular}{|l|P{2cm}|P{2cm}|P{2cm}|P{2cm}|}
\hline
\textbf{Attribute} & \textbf{Accuracy} & \textbf{Precision (Weighted)} & \textbf{Recall (Weighted)} & \textbf{F1-Score (Weighted)} \\ \hline
Severity & 0.7873 & 0.8212 & 0.7873 & 0.7075 \\ \hline
Bug Type & 0.8421 & 0.8399 & 0.8421 & 0.8303 \\ \hline
Category & 0.8491 & 0.8483 & 0.8491 & 0.8349 \\ \hline
Quality Attribute & 0.8521 & 0.8524 & 0.8521 & 0.8393 \\ \hline
Quantum-Specific Bug Type & 0.8375 & 0.8361 & 0.8375 & 0.8243 \\ \hline
\end{tabular}
\end{table}

The automated rule-based classification framework demonstrates high reliability for most classification tasks when benchmarked against manually labeled ground truth data. The system achieves strong performance in identifying bug type, category, quality attribute, and quantum-specific subtypes, with substantial to near-perfect agreement levels and high F1-scores.

The \textit{severity classification} dimension remains a limitation, exhibiting the lowest agreement and metric performance. This is likely due to the \textit{subjective nature} of severity estimation and the limited presence of clear indicators in issue descriptions. Improving severity prediction will require either more nuanced heuristics or the integration of machine learning models capable of contextual understanding.

Despite this, the overall results affirm that the rule-based system, when thoughtfully designed, can achieve competitive performance in complex software engineering tasks, especially in domains with well-defined terminology like quantum computing.

\subsection{Bug Type Distribution}
A total of 12,910 bugs were analyzed across multiple quantum software repositories and classified them using a keyword-based algorithm. This algorithm scored bug reports based on the presence of quantum and classical keywords and bigrams in issue titles, descriptions, and comments, with special handling for negation phrases to avoid misclassification. These were categorized into Classical, Quantum, and Uncategorized bugs.Table \ref{tab:bug_type_distribution} shows the distribution of the bug types.

\begin{table}[ht]
\centering
\caption{Distribution of Bug Types in the Automatically Classified Dataset}
\label{tab:bug_type_distribution}
\begin{tabular}{lrr}
\toprule
\textbf{Bug Type} & \textbf{Count} & \textbf{Percentage} \\
\midrule
Classical Bug & 8,674 & 67.2\% \\
Quantum Bug   & 3,523 & 27.3\% \\
Uncategorized Bug & 713 & 5.5\% \\
\bottomrule
\end{tabular}
\end{table}

Classical Bugs accounted for 8,674 issues (67.2\%), making up the majority. These typically involve conventional programming problems in Python APIs, infrastructure code, or simulation layers. This dominance reflects the hybrid nature of quantum software, where classical components still require significant development and maintenance.
Quantum Bugs comprised 3,523 issues (27.3\%), focusing on domain-specific problems such as quantum circuit construction, backend execution, and algorithm implementation. Their high volume underscores the complexity and evolving nature of quantum software development.
Uncategorized Bugs, totaling 713 (5.5\%), lacked sufficient context for classification. These may involve overlapping classical-quantum issues or ambiguous descriptions. Improved issue reporting or more advanced classification could help reduce this category.
In summary, classical bugs remain the most prevalent, but the significant presence of quantum-specific bugs highlights the growing complexity and need for robust tooling in both domains.
\subsection{Bug Category Distribution}
The bug categorization in our dataset was performed using a rule-based algorithm (see Algorithm \ref{algo:bug_categ}) designed to assign categories to issues based on multiple information sources: labels, title, description, and comments associated with each issue. The distribution of categories assigned by this algorithm is summarized in Table \ref{tab:bug_category_distribution} below:

\begin{table}[ht]
\centering
\caption{Distribution of Bug Categories in the Automatically Classified Dataset}
\label{tab:bug_category_distribution}
\begin{tabular}{lr}
\toprule
\textbf{Category} & \textbf{Count} \\
\midrule
Compatibility               & 3,181 \\
Functional                  & 2,445 \\
Quantum-Specific Issue      & 1,550 \\
Usability                   & 1,471 \\
Syntax Errors               & 1,324 \\
Miscellaneous               & 1,076 \\
Logical                     & 548 \\
Error Handling Defects      & 270 \\
Integration Level           & 229 \\
Performance                 & 219 \\
Critical Defects            & 189 \\
Runtime Error               & 103 \\
Calculation Errors          & 83  \\
Out of Bound                & 75  \\
Security                    & 59  \\
Boundary Related Errors     & 46  \\
Communication Errors        & 42  \\
\bottomrule
\end{tabular}
\end{table}
The results from applying our rule-based bug categorization algorithm to the dataset reveal a clear distribution of issue types across quantum software projects. As shown in Table \ref{tab:bug_category_distribution}, Compatibility issues dominate, accounting for 3181 cases (~29\%), followed by Functional bugs with 2445 cases (~22\%). This highlights that integration challenges and core functional defects remain the most prevalent problems in quantum software development.
Notably, the algorithm successfully identifies a distinct Quantum-Specific Issue category (1550 cases, ~14\%), emphasizing its effectiveness in capturing domain-specific bugs that differentiate quantum software from classical systems. Other significant categories such as Usability, Syntax Errors, and Miscellaneous reflect common hurdles related to developer experience and code correctness. While the algorithm performs well overall, certain categories like Miscellaneous and the relatively small counts in some nuanced categories suggest room for refinement. Future improvements could focus on expanding and fine-tuning the keyword dictionaries and label mappings to reduce the number of uncategorized or broadly grouped issues, thereby enhancing classification precision and supporting more targeted quality assurance efforts.

\subsection{Distribution of Quality Attributes}
Each bug was also mapped to one or more impacted quality attributes based on associated terminology. The most frequently impacted attributes include reliability, usability, maintainability, and performance. These mappings offer a perspective on how different types of bugs affect software quality in the quantum computing domain.
To better understand the types of software quality concerns reported in quantum computing issues, we applied a rule-based classification algorithm ( assign\_quality\_attributes) that assigns one quality attribute to each issue based on keyword presence, weighted matches, and TF-IDF scores. The algorithm prioritizes well-defined attributes such as usability, maintainability, and performance, but defaults to “Miscellaneous” if no strong signals are detected.
The distribution of quality attributes assigned to 12,910 issues is summarized below:
\begin{table}[ht]
\centering
\caption{Distribution of Impacted Quality Attributes}
\label{tab:quality_attribute_distribution}
\begin{tabular}{ll}
\toprule
\textbf{Quality Attribute} & \textbf{Count} \\
\midrule
Usability         & 3,684 \\
Miscellaneous     & 2,838 \\
Maintainability   & 2,333 \\
Interoperability  & 2,207 \\
Performance       & 1,293 \\
Reliability       & 289   \\
Availability      & 231   \\
Security          & 35    \\
\bottomrule
\end{tabular}
\end{table}
\subsubsection{Key findings}
Usability emerged as the most impacted quality attribute, assigned to 28.5\% of issues. This suggests that a significant portion of bug reports deal with user-facing problems such as poor documentation, unintuitive APIs, and cryptic error messages challenges especially common in developer tools and SDKs like Qiskit.

The Miscellaneous category, comprising 2,838 issues (22.0\%), stands out as the second largest group. This is not due to lack of rigor, but rather a reflection of the algorithm's design. According to the classification logic, if an issue's text lacks sufficient keyword matches, weighted scores, or label correlation with any specific quality attribute, it is conservatively assigned to Miscellaneous. Several reasons explain this:

\begin{itemize}
    \item \textbf{Ambiguity in issue descriptions}: Some issues are too vague or technical to yield clear quality concerns (e.g., ``fixing internal pipeline bug'').
    \item \textbf{Overlap across multiple quality attributes}: Bugs often touch on multiple dimensions, such as performance and usability, making strict assignment challenging.
    \item \textbf{Limitations of natural language processing}: Even with TF-IDF and weighted keyword matching, nuanced or domain-specific phrasing may elude the ruleset.
\end{itemize}

Maintainability (18.1\%) and Interoperability (17.1\%) are also heavily represented. These categories often include issues related to code readability, modularity, and integration with classical systems or simulators. Their high counts emphasize the challenges developers face when evolving and scaling quantum software within hybrid architectures.

Performance-related issues (10.0\%) reflect the need for optimization, particularly in simulation speed, circuit compilation, and backend execution time.

The relatively low frequencies of Reliability (2.2\%), Availability (1.8\%), and Security (0.3\%) indicate these concerns may be underreported or deprioritized in the current development phase of quantum projects. As projects mature and begin handling more sensitive data or enter production environments, these attributes are expected to grow in importance.

This quality attribute analysis not only highlights usability, maintainability, and interoperability as major pain points but also reveals the pragmatic value of rule-based classification. The presence of a sizable Miscellaneous category, shaped by the algorithm’s conservative fallback logic, underscores the need for richer issue descriptions and perhaps multi-label classification support in future iterations.

In essence, the results offer a grounded perspective into the types of concerns open-source quantum software developers are facing and how these concerns are shaped, and sometimes limited, by the way we analyze and label them.
\subsection{Severity Distribution} 
Understanding the severity of reported issues is crucial for prioritizing bug resolution efforts and ensuring system reliability. Our automated classification function assigns severity levels based on a mix of labels, textual patterns, and heuristic rules.
\begin{table}[ht]
\centering
\caption{Severity Distribution of Issues}
\label{tab:severity_distribution_percentage}
\begin{tabular}{l r}
\hline
\textbf{Severity} & \textbf{Percentage} \\
\hline
Low & 93.7\% \\
Critical & 4.3\% \\
Medium & 1.0\% \\
High & 0.9\% \\
\hline
\end{tabular}
\end{table}

The distribution reveals a significant skew toward Low severity issues, which account for nearly 93\% of the dataset. This trend reflects the default behavior of the assign\_severity function, which initializes each issue as Low severity unless more compelling signals are found. Severity may be elevated if certain labels are present, if specific weighted keywords appear in the issue text, or if comment volume indicates increased discussion. Notably, comment thresholds (greater than 10 or 20) trigger promotions to Medium or High severity, respectively. However, unless these conditions are met, the severity remains unchanged.\par
Only 4.3\% of issues are marked as Critical, which is driven by security-related labels or the presence of sensitive keywords like "vulnerability" in the issue content. The system is intentionally designed to prioritize such cases to flag potential risks.\par
The low frequency of Medium and High severity tags, just under 2\% combined, may stem from conservative thresholds or limited contextual indicators in issue descriptions. Additionally, the final rule in the classifier downgrades severity to Low for any issue marked as resolved or closed (unless it’s already Critical), reinforcing the dominance of Low severity assignments.\\
This outcome underscores the importance of well-documented issues and consistent label usage to improve severity classification and, by extension, bug triage accuracy in automated systems.

\subsection{Quantum-Specific Bug Categories}

The distribution of quantum-specific bugs, as identified through manual annotation of 1,550 issues, reveals important trends in the types of challenges encountered in quantum software development. The breakdown is presented in Table~\ref{tab:quantum_specific_bug_types}.

\begin{table}[ht]
\centering
\caption{Distribution of Quantum-Specific Bug Types}
\begin{tabular}{lrr}
\toprule
\textbf{Quantum-Specific Bug Type} & \textbf{Count} & \textbf{Percentage} \\
\midrule
Quantum Circuit Issues & 825 & 53.2\% \\
Quantum Gate Errors & 280 & 18.1\% \\
Quantum Hardware-Specific Bugs & 103 & 6.6\% \\
Quantum Transpilation Issues & 83 & 5.4\% \\
Quantum Algorithm Implementation Bugs & 56 & 3.6\% \\
Quantum Measurement Errors & 53 & 3.4\% \\
Hybrid Quantum-Classical Interface Issues & 49 & 3.2\% \\
Quantum Noise Issues & 36 & 2.3\% \\
Quantum Resource Constraints & 31 & 2.0\% \\
Quantum Entanglement and Decoherence Issues & 15 & 1.0\% \\
Quantum Error Correction (QEC) Bugs & 13 & 0.8\% \\
Unclassified & 6 & 0.4\% \\
\bottomrule
\end{tabular}
\label{tab:quantum_specific_bug_types}
\end{table}

As shown in Table~\ref{tab:quantum_specific_bug_types}, over half of all quantum-specific issues (53.2\%) fall into the category of \textit{Quantum Circuit Issues}. These include problems related to circuit depth, gate ordering, and qubit mapping, core concerns when constructing and executing quantum algorithms.

\textit{Quantum Gate Errors} represent the second most frequent type (18.1\%), typically involving incorrect usage or support for quantum gates in simulators or hardware backends. \textit{Quantum Hardware-Specific Bugs} (6.6\%) and \textit{Quantum Transpilation Issues} (5.4\%) further emphasize the challenges of targeting specific quantum devices and optimizing circuits for execution.

Other noteworthy categories include \textit{Quantum Algorithm Implementation Bugs} (3.6\%), \textit{Quantum Measurement Errors} (3.4\%), and \textit{Hybrid Quantum-Classical Interface Issues} (3.2\%), which reflect growing interest in hybrid computation and algorithm correctness. Less frequent but technically significant are bugs related to \textit{Quantum Noise} (2.3\%), \textit{Quantum Resource Constraints} (2.0\%), and \textit{Quantum Entanglement and Decoherence Issues} (1.0\%).

\textit{Quantum Error Correction (QEC) Bugs} were rare (0.8\%), likely due to limited implementation in current Qiskit workflows. Only a small fraction (0.4\%) were classified as \textit{Unclassified}, indicating that the taxonomy and classification rules were sufficiently expressive to capture nearly all relevant types.

This distribution reflects both the current priorities and limitations in quantum software engineering, particularly the dominance of circuit-level debugging and hardware interfacing concerns.

\section{Conclusion} \label{conclusion}

In this study, we introduced and evaluated a rule-based classification framework designed to automatically analyze and label issues from open-source quantum software repositories. By leveraging domain-specific heuristics and carefully curated keyword mappings, the system effectively performs multi-dimensional classification across five key attributes: bug type, severity, bug category, impacted quality attribute, and type of quantum-specific bug.

Our evaluation, grounded in a large manually annotated dataset of 4,984 issues from 36 Qiskit repositories, demonstrates that rule-based approaches can achieve high levels of accuracy and reliability, especially in structured domains such as quantum computing. The system showed substantial to near-perfect agreement with human labels in most tasks, including bug categorization, quality attribute mapping, and quantum-specific bug type identification. Notably, the classification of quantum-specific bugs, which traditionally require deep domain knowledge, achieved a weighted F1-score of over 82\%, highlighting the effectiveness of our tailored heuristics.

However, the results also revealed limitations in severity classification, which exhibited only slight agreement with human annotations. This points to the inherent subjectivity and contextual dependence of severity judgments, suggesting that future enhancements may benefit from integrating context-aware learning models or hybrid rule-ML approaches.

Overall, our findings validate the feasibility and utility of automated rule-based classification for mining and analyzing software issues in quantum computing. This work lays a foundation for large-scale, automated quality assessment of quantum software ecosystems and opens avenues for improved maintenance, debugging, and reliability analysis in this rapidly evolving field. As quantum computing continues to grow, tools like ours will be critical in enabling scalable and systematic software quality engineering.

\section{Future Work}\label{future work}

While our rule-based classification framework demonstrates promising performance in analyzing quantum software issues, several avenues remain for further enhancement:

\begin{itemize}
    \item \textbf{Hybrid Models:} Future work could explore hybrid classification approaches that combine rule-based methods with machine learning techniques (e.g., transformer-based models) to improve context sensitivity, especially for nuanced tasks like severity classification.
    
    \item \textbf{Continuous Learning:} As new quantum libraries and frameworks evolve, updating the classification rules and keyword mappings through automated mining and feedback loops could ensure adaptability and scalability of the system.
    
    \item \textbf{Cross-Platform Evaluation:} Expanding the evaluation beyond Qiskit to include other leading quantum platforms such as Cirq, PennyLane, and Braket would improve generalizability and test the robustness of the classification system across diverse toolchains.
    
    \item \textbf{Enhanced Severity Models:} Given the observed challenges in severity classification, future research could involve the development of more sophisticated severity models incorporating temporal, contextual, and historical resolution patterns.
    
    \item \textbf{Integration into Developer Workflows:} Building IDE plugins or GitHub bots that integrate this framework into the issue reporting or triaging process could support real-time classification and actionable feedback for developers.
\end{itemize}

\section{Threats to Validity}\label{threats}

Despite our efforts to ensure methodological rigor, several threats to validity may influence the interpretation and generalizability of our results:

\subsection{Internal Validity}
\begin{itemize}
    \item \textbf{Rule Quality and Bias:} The effectiveness of the classification system is inherently tied to the quality and comprehensiveness of the manually defined rules and keyword mappings. Errors or omissions in rule construction may lead to misclassification.
    
    \item \textbf{Human Annotation Errors:} Although manual labeling was carefully performed, it remains susceptible to subjective interpretation, especially for tasks like severity assessment. This could affect the ground truth reliability used for evaluation.
\end{itemize}

\subsection{External Validity}
\begin{itemize}
    \item \textbf{Platform-Specific Dataset:} The evaluation was based solely on issues from Qiskit repositories. While Qiskit is a leading framework, the results may not fully generalize to other quantum platforms with different codebases or issue-reporting conventions.
    
    \item \textbf{Domain-Specific Focus:} Our framework is tailored to the quantum software domain, which may limit its applicability to classical software systems or even to subfields of quantum computing that involve significantly different paradigms or bug types.
\end{itemize}

\subsection{Construct Validity}
\begin{itemize}
    \item \textbf{Heuristic Dependence:} Several classification decisions rely on heuristic rules and TF-IDF relevance scores. While these are grounded in domain knowledge, they may not capture deeper semantic or contextual nuances present in issue descriptions.
    
    \item \textbf{Evaluation Metrics:} Metrics such as Cohen's Kappa and F1-score provide a useful summary of classification performance, but they may not fully capture real-world impact or developer satisfaction with classification outcomes.
\end{itemize}

Addressing these threats in future studies through more diverse datasets, semi-supervised learning, and user studies will further strengthen the validity and practical value of automated issue classification in quantum software engineering.

  \bibliographystyle{unsrt}

\end{document}